\documentclass{emulateapj}
\usepackage{subfigure,epsfig,url}
\shorttitle{Inner Oort Cloud LPC}
\shortauthors{Kaib et al.}

\begin{document}

\title{2006 SQ$_{372}$: A Likely Long-Period Comet from the Inner Oort Cloud}

\author{Nathan A. Kaib\altaffilmark{1}, Andrew C. Becker\altaffilmark{1}, R. Lynne Jones\altaffilmark{1}, Andrew W. Puckett\altaffilmark{2}, Dmitry Bizyaev\altaffilmark{3}, Benjamin Dilday\altaffilmark{4,5,9}, Joshua A. Frieman\altaffilmark{5,6,7}, Daniel J. Oravetz\altaffilmark{3}, Kaike Pan\altaffilmark{3}, Thomas Quinn\altaffilmark{1}, Donald P. Schneider\altaffilmark{8}, and Shannon Watters\altaffilmark{3}}
\altaffiltext{1}{Department of Astronomy, University of Washington, Box 351580, Seattle, WA 98195-1580, kaib@astro.washington.edu.}
\altaffiltext{2}{Department of Physics and Astronomy, University of Alaska Anchorage, 3211 Providence Dr., Anchorage, AK 99508}
\altaffiltext{3}{Apache Point Observatory, P.O. Box 59, Sunspot, NM 88349-0059}
\altaffiltext{4}{Department of Physics, University of Chicago, Chicago, IL 60637}
\altaffiltext{5}{Kavli Institute for Cosmological Physics, University of Chicago, Chicago, IL 60637}
\altaffiltext{6}{Center for Particle Astrophysics, Fermi National Accelerator Laboratory, P.O. Box 500, Batavia, IL 60510}
\altaffiltext{7}{Department of Astronomy and Astrophysics, The University of Chicago, 5640 South Ellis Avenue, Chicago, IL 60637}
\altaffiltext{8}{Department of Astronomy and Astrophysics, Pennsylvania State University, University Park, PA 16802}
\altaffiltext{9}{Department of Physics and Astronomy, Rutgers, the State University of New Jersey, 136 Frelinghuysen Rd, Piscataway, NJ 08854-0819}

\begin{abstract}
We report the discovery of a minor planet (2006 SQ$_{372}$) on an orbit with a perihelion of 24 AU and a semimajor axis of 796 AU.  Dynamical simulations show that this is a transient orbit and is unstable on a timescale of $\sim$200 Myrs.  Falling near the upper semimajor axis range of the scattered disk and the lower semimajor axis range of the Oort Cloud, previous membership in either class is possible.  By modeling the production of similar orbits from the Oort Cloud as well as from the scattered disk, we find that the Oort Cloud produces 16 times as many objects on SQ$_{372}$-like orbits as the scattered disk.  Given this result, we believe this to be the most distant long-period comet ever discovered.  Furthermore, our simulation results also indicate that 2000 OO$_{67}$ has had a similar dynamical history.  Unaffected by the ``Jupiter-Saturn Barrier,'' these two objects are most likely long-period comets from the inner Oort Cloud.
\end{abstract}


\section{Introduction}

In the past three decades, many minor planets have been observed on planet-crossing orbits in the outer solar system.  These objects are transient in nature \citep{tismal03,hol93} and, therefore, must be continuously resupplied from more stable reservoirs.  Previous numerical modeling has shown that both the scattered disk and Oort Cloud contain large populations of bodies diffusing into the planetary region on Gyr timescales at a rate that can account for the planet-crossing populations observed today \citep{levdun97,wietre99,emel07}.  In the scattered disk, trans-Neptunian objects (TNOs) orbit the Sun on eccentric orbits with perihelia taking them to within 10 AU of Neptune or less.  Driven by the effects of Neptunian perturbations, these orbits undergo a chaotic evolution over time \citep{dunlev97}.  Due to this process, some of these bodies eventually begin to scatter more strongly off of Neptune, and because complete ejection by Neptunian encounters is quite unlikely, the perihelia of these orbits tend to evolve further inward \citep{fern80a}.  

While the scattered disk is typically assumed to be the source of most observed planet-crossers in the outer Solar System \citep{dun04}, there is reason to expect a substantial Oort Cloud contribution as well, especially at high semimajor axes \citep{emel05}.  Oort Cloud orbits are continually driven into the planetary region via perihelion shifts due to the external perturbations of the Galactic tide and passing stars \citep{oort50,fern80,heistre86}.  When Oort Cloud bodies have their perihelia pushed back into the planetary region, their orbital evolution is largely determined by their semimajor axis \citep{hills81}.  For low semimajor axis orbits ($a \lesssim$ 20,000 AU), perihelion migration through the planetary region occurs very slowly over many orbital periods because the external perturbations driving this migration are small compared to the binding force of the Sun.  Thus, as these objects slowly drift Sunward, they will inevitably encounter Jupiter and Saturn, receiving a gravitational energy kick that ejects these weakly bound bodies to interstellar space.  On the other hand, the perihelia of high-semimajor axis Oort Cloud objects ($a \gtrsim$ 20,000 AU) evolve much more rapidly, and these bodies' perihelia can move from outside 15 AU to inside 5 AU in less than one orbital period.  As a result, before being ejected by Jupiter or Saturn, these objects can pass through the terrestrial planet region where we subsequently detect them as long-period comets (LPCs).  

Thus, the Oort Cloud can be divided into an outer Oort Cloud ($a >$ 20,000 AU), from which all known LPCs originate, and an inner Oort Cloud ($a <$ 20,000 AU) that is not sampled by known LPCs but is theorized to exist due to models of the comet cloud's formation \citep{dun87}.  Because of this, obtaining a sample of comets that probes the entire Oort Cloud requires detecting LPCs outside the gravitational barriers of Saturn and Jupiter ($q \gtrsim 15$ AU).  Due to the steep decrease in brightness with heliocentric distance, however, detection of these objects is very challenging.  Until this work only one LPC, Comet C/2003 A2 ($q = 11.4$ AU), has been observed with a perihelion beyond Saturn's orbit \citep{gleas03,green03}.  Detecting LPCs beyond $\sim$15 AU would test Oort Cloud formation models \citep{dun87,dones04,emel07,kaibquinn08} as well as constrain the birthplace environment of the Sun \citep{fern97,fernbrun00,bras06,kaibquinn08}.  For our purposes, we define an LPC as any Oort Cloud body that has been reinjected into the planetary region.  These objects may or may not show cometary activity.

Here we report a likely LPC well outside the Jupiter-Saturn barrier with $q = 24.2$ AU.  Section 2 discusses our observations of this object.  Section 3 is devoted to modeling the production of similar objects from both the Oort Cloud and the trans-Neptunian scattered disk.  In Section 4, we use our simulations to make the case that 2006 SQ$_{372}$ is indeed from the inner Oort Cloud rather than the scattered disk.

\section{Observations}

2006 SQ$_{372}$ was initially discovered in images taken by the SDSS--II Supernova Survey (Frieman et al. 2008; see also Fukugita et al. 1996, Gunn et al. 1998, Gunn et al. 2006, York et al. 2000).  This survey repeatedly scanned a 300 sq. deg. region (SDSS Stripe 82) centered on the celestial equator in the southern Galactic hemisphere for three months out of the year (Sept - Nov) from 2005--2007.  This area also passes through the Solar System ecliptic plane, spanning ecliptic latitude $-20\deg < \beta < 20\deg$ and ecliptic longitude $-58\deg < \lambda < 58\deg$.  A search through these data by Becker et al. (In preparation) attempted to link all single--epoch transients into orbits consistent with Keplerian motion around the Sun, using the linking software of \cite{2007Icar..189..151K} and orbit fitting software of \cite{2000AJ....120.3323B}.  Nearly 50 trans--Neptunian objects were discovered in this effort, including 2006 SQ$_{372}$.

2006 SQ$_{372}$ was first discovered as a single--epoch transient on the night of Sept 27, 2006 at Right Ascension 21:09:20.3 and Declination +01:09:36 (J2000), at an apparent magnitude of r = 21.59 $\pm$ 0.06 in the magnitude system of \cite{1999AJ....118.1406L}.  The data are astrometrically calibrated using the methods described in \cite{2003AJ....125.1559P} and have been photometrically recalibrated using the methods described in \cite{2007AJ....134..973I}.  Subsequent observations throughout the 2006 observing season recovered this object on 9 additional dates as it moved through Stripe 82.  These observations were used to calculate an orbital ephemeris, and to precover additional data from 10 nights in 2005, as well as 7 subsequent nights in 2007.  The orbital ephemeris is available through the minor planet data center, and detailed observational data can be provided on request and is also posted at http://www.boulder.swri.edu/~buie/kbo/astrom/ 06SQ372.html.

The current orbital arc now spans 2.9 years.  Using the Jet Propulsion Laboratory (JPL) {\tt Horizons} software\footnote{http://ssd.jpl.nasa.gov/?horizons}, the barycentric orbital elements for this system at epoch 2008 May 14.0 are semi--major axis $a = 796 \pm 3$ AU, eccentricity $e = 0.9696 \pm 0.0001$, and inclination $i = 19.46471 \pm 0.00008$ degrees.  For comparison, the Minor Planet Center (MPC) software provides a best-fit orbit of $a/e/i$ = $1057/0.9771/19.46052$ evaluated in heliocentric coordinates.  For objects at such high eccentricity, barycentric coordinates are more stable than heliocentric coordinates, and orbital elements are evaluated with respect to the Solar System barycenter for the remainder of this work.  2006 SQ$_{372}$ reached its perihelion of $24.1667 \pm 0.0004$ AU on 2006 Aug 12, and has an orbital period of 22,466 years.  

In Figure 1, we compare the perihelion and semimajor axis of 2006 SQ$_{372}$ to all other TNOs and Centaurs listed in the MPC.  We see that the semimajor axis of this object is larger than any other known body with perihelion in the outer Solar System.  The next two largest semimajor axes are those of 2000 OO$_{67}$ and Sedna.  While Sedna's large perihelion makes it dynamically isolated from the outer planets \citep{morblev04}, 2000 OO$_{67}$ is in a Neptune-crossing orbit, making it most similar to 2006 SQ$_{372}$.

\begin{figure}[htbp]
\centering
\includegraphics[scale=.5]{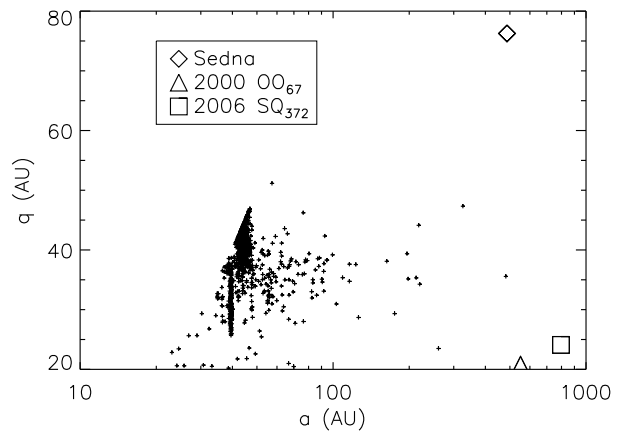}
\caption{Plot of $q$ vs. $a$ for all listed TNOs and Centaurs in the Minor Planet Center.  The orbits of Sedna, 2000 OO$_{67}$, and 2006 SQ$_{372}$ are shown with diamond, triangular, and square data points respectively.  All other observed objects are displayed with smaller crosses.}\label{fig:1}
\end{figure}

\section{Numerical Methods}

All numerical work in this paper is done using the SWIFT RMVS3 integration package \citep{levdun94} modified as in \citet{kaibquinn08}, and includes the gravitational effects of the Sun, giant planets, passing stars \citep{rick08}, and Galactic tide \citep{lev01}.  We perform three different simulations to investigate the dynamics of 2006 SQ$_{372}$.  The first is a simulation of 10$^3$ test particles cloned from the observed orbit of 2006 SQ$_{372}$ to determine how this orbit will evolve in the future.  This first simulation motivates our other numerical work, which we use to constrain the dynamical origin of this object.  To determine the most likely origin scenario for 2006 SQ$_{372}$, we run two different simulations: one modeling the production of similar bodies from the scattered disk and the other from the Oort Cloud.  It must be noted that this approach assumes that the scattered disk and the Oort Cloud are the only two populations capable of producing planet-crossing orbits in the outer solar system.  While previous work has shown that classical KBOs and resonant TNOs can also evolve on to Neptune-crossing orbits \citep{dunlev97,morb97}, the occurence of this is substantially less frequent than in the scattered disk \citep{dun04}.  Our three different simulations are described in the following subsections.  

\subsection{Preliminary Clone Simulations}

With a perihelion inside Neptune's orbit, the orbit of 2006 SQ$_{372}$ appears to be transient in nature.  To verify this we first integrate 10$^3$ clones of this object.  These clone orbits are generated from a multivariate distribution using the covariances between orbital elements.  We find that the population of clones in planet-crossing orbits has a half-life of $\sim$180 Myrs.  After 1 Gyr, only 2\% of our clones are still found in Neptune-crossing orbits, while 62\% have been ejected from the solar system ($r >$ 200,000 AU), and 36\% reside in the Oort Cloud ($q >$ 45 AU).  

Because 2006 SQ$_{372}$ is on a planet-crossing orbit not stable over the lifetime of the Solar System, it must have previously belonged to another class of objects.  We believe the Oort Cloud and scattered disk are the two most likely source populations for 2006 SQ$_{372}$, as both of these efficiently place bodies on planet-crossing orbits \citep{wietre99,levdun97}.  An Oort Cloud origin would imply that torques from passing stars and the Galactic tide reinjected its perihelion into the planetary region after which planetary encounters decreased its semimajor axis.  However, it is also possible that 2006 SQ$_{372}$ evolved from the scattered disk.  In this scenario, the perihelion of 2006 SQ$_{372}$ chaotically diffused inward before its semimajor axis was inflated via planetary encounters.

\subsection{Scattered Disk Production}

To determine the likelihood that SQ$_{372}$ is from the scattered disk, we simulate the production of similar orbits from this population.  To do this, we replicate the work of \citet{disistbrun07} who modeled the production of Centaurs from orbits cloned from observed scattered disk objects.  Particles are removed from our simulation if they reach $r >$ 6,000 AU or collide with a giant planet or the Sun.

We begin our simulation with 2,500 test particles and the four giant planets on their present orbits and integrate for 4.5 Gyrs.  The initial conditions we use to assign test particle orbits are very similar to \citet{disistbrun07} and \citet{fern04} and are based on the orbital distribution of observed SDOs.  To first obtain a sample of real SDO orbits, we classify any observed TNO with 30 AU $< q <$ 40 AU and $a >$ 50 AU as an SDO.  Additionally, we also include observed TNOs with 30 AU $< q <$ 40 AU, 40 AU $< a <$ 50 AU, and $e >$ 0.2.  This selected set of real orbits then forms the basis for our initial perihelion distribution, which we generate as follows.  First, we weight the perihelion distribution of real orbits by the fraction of orbital period each object spends inside 45 AU, an approximate zone of observability.  Using this weighted distribution, we then assign perihelia to our test particles.  Next we assign semimajor axes between 40 AU and 200 AU to particles by randomly drawing from a distribution proportional to $a^{-2}$.  We choose this power-law distribution because \citet{disistbrun07} found it to be a good fit to their debiased distribution of real SDO semimajor axes.  To assign orbital inclinations, we use the observed SDO inclination distribution obtained in \citet{brown01} of 
\begin{equation}
F\left(i\right)\propto \sin{i}\exp{-i^{2}/2\sigma_{i}^{2}}
\end{equation}
where we choose $\sigma_{i} = 12^{\circ}$ as in \citet{disistbrun07}.  Lastly, arguments of perihelion, longitudes of ascending node, and mean anomalies are assigned randomly.    

Because we are interested in the distribution of orbits penetrating the planetary region from the scattered disk, we clone a particle 20 times if it evolves to an orbit with $q <$ 30 AU during the course of our simulation.  This cloning is done by shifting each cartesian coordinate by a random number between $\pm$1 x 10$^{-6}$ AU.  These clones are then evolved until collision with the Sun or a giant planet, or ejection ($r >$ 6,000 AU).  

We choose to remove particles with $r >$ 6,000 AU because we do not want to include any SDOs that move to the Oort Cloud and then subsequently evolve to planet-crossing orbits from there.  However, doing this cuts off a possible dynamical pathway where $a$ temporarily random walks above 3,000 AU and is subsquently pulled back down to SQ$_{372}$ values while $q$ never leaves the planetary region.  To evaluate the importance of this pathway, we continue to integrate a subset of the particles removed at $r =$ 6,000 AU.  Based on these integrations, we find that SQ$_{372}$-like orbits are attained about an order of magnitude less frequently than for particles yet to reach $r >$ 6,000 AU.  Thus, our simulation does capture the dominant mechanism producing 2006 SQ$_{372}$ analogues from the scattered disk.

\subsection{Oort Cloud Production}

To evaluate the rate the Oort Cloud produces orbits similar to 2006 SQ$_{372}$, we dynamically evolve 10$^6$ Oort Cloud particles for 1.4 Gyrs under the influence of the giant planets, Galactic tide, and passing field stars.  To simulate this large number of particles efficiently, we have used the variable timestepping procedure outlined in \citet{kaibquinn08} with an extra level of larger timesteps.  The base timesteps are $t = $0.25, 10, and 50 yrs for $r < 300$ AU, 300 AU $< r <$ 400 AU, and $r >$ 400 AU respectively.  

To choose our initial particle orbits, we use orbital element distributions from the $\sim10^3$-particle Oort Cloud that exists at the end of the 4.5 Gyr large control simulation of \citet{kaibquinn08}.  First, we assign semimajor axes by randomly drawing from the $a$ cumulative distribution function (CDF) from our formation simulation.  In this formation simulation $a$ ranges from 911 AU to $\sim$10$^5$ AU.  Next we divide our formation simulation into five separate ranges of $a$, each containing an equal number of particles.  For each $a$-range, we then construct CDFs for $J/L$, $J_z/J$, $\Omega_G$, and $\omega_G$, where $L$ is the first Delaunay momentum ($\sqrt{GM_\Sun a}$), $J$ is the total angular momentum, $J_z$ is the z-component angular momentum in the galactic frame, $\Omega_G$ is the longitude of ascending node in the galactic frame, and $\omega_G$ is the argument of perihelion in the galactic frame.  These CDFs are then used to randomly assign $e$, $i$, $\Omega$, and $\omega$ for the new initial orbits in each $a$-range.  Finally, mean anomalies are assigned randomly from a uniform distribution.  

The goal of this procedure is to build a $10^6$-particle Oort Cloud that looks very similar to the $\sim10^3$-particle cloud formed in \citet{kaibquinn08}.  This is of particular importance in the inner Oort Cloud where orbits are less isotropized.  Due to the very long Kozai cycle times for the galactic tide inside $a \lesssim$ 5,000 AU, the extreme inner Oort Cloud should contain significantly more orbits with $\dot{q} >$ 0 than with $\dot{q} <$ 0 \citep{koz62,heistre86}.  Failing to replicate this effect would enhance the rate that extreme inner Oort Cloud objects enter the planetary region and could artficially increase the production of orbits similar to 2006 SQ$_{372}$.  \citet{heistre86} show that $\dot{q} \propto -\sqrt{q}a^2\sin{2\omega_G}$.  Because we apply a different $\omega_G$ distribution for each range of $a$, our $10^6$-particle simulation does capture this bias at small $a$.  Moreover, it is likely that much of the inner Oort Cloud was randomized by perturbations of an early star cluster environment not included in the \citet{kaibquinn08} control simulation, and this would greatly diminish this $\dot{q}$ bias \citep{bras08}.  For this last reason, we can assume that the initial conditions for our simulation are conservative in this respect.

Because planet-crossing orbits typically result in ejection from the Solar System, there should be a constant deficiency of these orbits within the Oort Cloud.  However, the orbital distributions from the \citet{kaibquinn08} Oort Cloud formation simulation are too coarse to resolve this deficiency.  As a result, this part of the Solar System is initially overpopulated with Oort Cloud orbits.  To correct for this effect and allow the planets to eject these bodies, we evolve our comets under the gravitational influence of the Sun, four giant planets, passing stars, and the Milky Way tide for $\sim$1 Gyr.  In the final 400 Myrs of our simulation, external perturbations rather than our initial conditions control the population of planet-crossing orbits at all Oort Cloud semimajor axes.

\section{Simulation Results and Discussion}

\subsection{Comparison with Previous Work}

To confirm that our simulations model the dynamics of the scattered disk and Oort Cloud properly, we compare our generated dynamical populations with past work.  First, in the case of the scattered disk population, we replicate the analysis of \citet{disistbrun07} who model Centaur production from SDOs and obtain a time-weighted perihelion distribution for our Centaur population.  Upon comparing the two perihelion datasets, we find that a K-S test gives a 97\% probability that they were drawn from the same distribution.  Furthermore, we can also examine the mean lifetime of Centaurs in our simulation.  When we apply the $q < 5.2$ AU and $a > 1000$ AU removal rules used in \citet{disistbrun07} to our data, we obtain a mean Centaur lifetime of 77 Myrs, which is comparable to their value of 72 Myrs.

In the case of our Oort Cloud model, we compare our simulation to that of \citet{emel07} who modeled the production of comets from the Oort Cloud.  By measuring the rate that dynamically new comets enter the inner 5 AU of the solar system in their simulations and assuming a real rate of 10 yr$^{-1}$, \citet{emel07} calculate that there must be 4.5--5.5 x 10$^{11}$ bodies in the Oort Cloud greater than 1 km in diameter.  We perform an analogous calculation with our Oort Cloud simulations and determine that the Oort Cloud must contain $\sim$5 x 10$^{11}$ bodies, which is in excellent agreement.  Even more encouraging is that the simulations of \citet{emel07} predict objects similar to 2006 SQ$_{372}$ passing through the outer solar system from the Oort Cloud.

\subsection{2006 SQ$_{372}$ Production Rates}

The goal of our numerical work is to determine the relative rates that 2006 SQ$_{372}$ analogues are produced from the Oort Cloud and scattered disk.  To do this, we sample the orbital elements of all particles at discrete timesteps.  For the scattered disk simulation, we sample our cloned orbits every 10$^{5}$ years, and our Oort Cloud simulation is sampled every 10$^{6}$ years.  To minimize the effects of our initial conditions, we only use the latter portions of each simulation.  For the scattered disk, we analyze the last 3.5 Gyrs, whereas in the Oort Cloud simulation we only use the final 400 Myrs.  We also choose these time lengths because they yield roughly equal numbers of orbits with $q <$ 30 AU.  Using this time-sampled data, we then generate a two-dimensional histogram of orbits in $q$-$\log{a}$ space for both datasets.  The grid spacing we use is 2 AU for $q$ and 0.278 for $\log{a}$.  Tracing the particles' evolution as they traverse through our $q$-$\log{a}$ histogram then allows us to construct a residence time distribution for orbits leaking into the planetary region from each reservoir.  

Once we have compiled orbital residence time histograms for both of our simulations we must now find a way to normalize each histogram bin population to match our own solar system.  In the case of our scattered disk simulation, we can use the number of Jupiter Family Comets (JFCs) produced to normalize our histogram, whereas for our Oort Cloud simulation we can normalize our dataset to match the rate new LPCs are produced.  Because the observed JFC population is dominated by bodies greater than 1 km in diameter \citep{tan06}, we impose a similar minimum diameter ($D >$ 1 km, or $H_{10}<$ 11) on the observed flux of LPCs used in our normalization routine.  Unfortunately, numerous parameters used to generate these normalization factors are uncertain by up to an order of magnitude.  In the following, we explore a range of normalization factors for each simulation.  

We begin by discussing normalization our scattered disk simulation.  In our own Solar System, there are 250 known active JFCs (2 $< T_J <$ 3, $q <$ 2.5 AU).   Given their proximity to Earth and their bright absolute magnitudes, we can take this as an observationally complete sample.  In addition, however, there should be a second population of dormant JFCs.  Based on models of physical aging, \citet{jew04} estimates that there are about twice as many dormant JFCs as active ones.  In contrast, by modeling the diffusion of JFC inclinations with time, \citet{levdun97} argue for a 3.5:1 dormant to active JFC ratio and possibly as high as 7:1.  Lastly, \citet{fernmorb06} find a ratio between 1 and 3 by studying dormant JFCs found in NEO observations.  \citet{volkmal08}, who derive the size of the scattered disk population based on known JFCs, adopt a ratio of 2 because of the agreement between \citet{fernmorb06} and \citet{jew04}.  For the same reason, we choose this value as the most probable one as well.  This implies a total number of 750 JFCs in the current solar system.  Because we find 546 instances of JFC orbits in our time-sampled data, we multiply the population in each of our histogram bins for the scattered disk simulation by a factor of 750/546.

In the case of our Oort Cloud orbital residence time histogram, we use the production rate of km-sized LPCs to determine a normalization factor.  Unfortunately, the observed flux of new LPCs in the inner solar system is not well-constrained.  \citet{ever67a,ever67b} give a value of $\sim$10 dynamically new comets per year inside 5 AU, and this is used in many numerical works that model the Oort Cloud population e.g., \cite{heis90,weiss96,dones04,emel07}.  However, using only LPCs observed in the LINEAR survey (where observational biases are much better characterized), \citet{fran05} finds a shallower size distribution than \citet{ever67b} and argues for a flux of 4 new LPCs/yr instead.  Further analyzing the LINEAR LPC observations, \citet{nes07} finds an even lower new LPC flux due to different discovery efficiencies between new and returning LPCs.  Although \citet{nes07} does not directly specify an LPC flux, the paper states that the \citet{heis90} assumption of 2.1 new LPCs/yr inside 1 AU is likely an order of magnitude too large.  In our simulations, this would correspond to an observed flux estimate of 1.5 new LPCs/yr inside 5 AU.  To be conservative, we adopt this last flux estimate.  In our Oort Cloud simulation, we find that 9 dynamically new LPCs pass through the inner 5 AU of the solar system every Myr on average.  As a result, we multiply all of our Oort Cloud histogram bins by [1.5/(9 x 10$^{-6}$)]/400.  The factor of 400 is included because we have performed 400 different time samplings of orbital elements, and we are normalizing to a rate of comet production rather than a steady state population as in the case of JFCs.  

In addition to the uncertainties in normalizing our simulations to JFCs and LPCs, our results are also affected by the uncertainty in the inner Oort Cloud population size.  The reason the inner Oort Cloud population size is poorly known is because no LPCs near Earth come from this reservoir.  Because of this, its uncertainty has no impact on our overall dataset normalization, but it may have significant consequences on SQ$_{372}$ production rates since inner Oort Cloud bodies with $q \gtrsim 15$ AU are not deflected by the Jupiter-Saturn barrier.  The simulation we use is based on the \citet{kaibquinn08} control simulation, which predicts a 1.5:1 inner-to-outer Oort Cloud population ratio.  Although \citet{dones04} finds a slightly lower ratio of 1:1, most simulations modeling Oort Cloud formation predict higher ratios, with some as high as 10:1 \citep{bras06,emel07,kaibquinn08,bras08}.  Because the inner Oort Cloud population was likely enriched by strong perturbations of the Sun's birthplace environment, the \citet{kaibquinn08} control simulation probably underestimates the inner Oort Cloud population size.  Based on the results of \citet{kaibquinn08} and \citet{bras06,bras08}, we believe that a 3:1 inner-to-outer Oort Cloud population is reasonable.  To account for the full range of possible population ratios, we weight particles with an initial $a <$ 20,000 AU by a given factor to model more anemic or enriched inner Oort Clouds.  For instance, we weight these particles by a factor of 2 to model the orbital distribution of our chosen inner-to-outer Oort Cloud ratio of 3:1.

With both of our histogram datasets normalized to the current solar system and an inner Oort Cloud model chosen, we now have orbit distributions through the planetary region for bodies originating from both the scattered disk and Oort Cloud.  Next we divide the Oort Cloud histogram dataset by the scattered disk one.  The result gives the ratio of orbits produced from the Oort Cloud to the number produced from the scattered disk in each histogram bin.  Now for an orbit with a given $q$ and $a$ we have a means to estimate the probability that it is from the Oort Cloud vs. the scattered disk.  

Before proceeding further, it is necessary to determine how accurate this method of mapping orbit production is.  For instance, if a particle is improbably scattered into a relatively stable orbit, its orbit will be sampled over and over for the entire simulation.  As a result, our map will be artificially distorted to reflect a high production rate for that particular orbit, even though the chance of the original scattering may be quite small.  To guard against this effect, we attempt to simulate a large enough numbers of particles so that this type of error is diminished to an acceptable level.  To test whether we have achieved this, we subdivide our particles into 4 groups, treat each one as a separate simulation, and then compare the results.  We find that the different probability ratios for the orbital bins near 2006 SQ$_{372}$ have a standard deviation of 16\%.  We can take this standard deviation as a conservative estimate of our real error.

\begin{figure}[htbp]
\centering
\includegraphics[scale=.5]{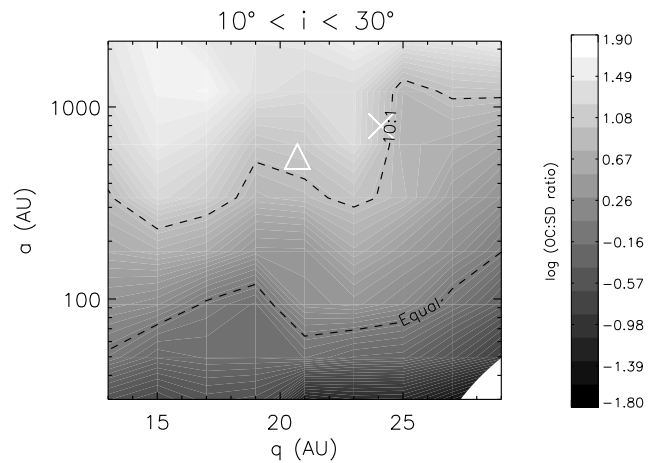}
\caption{Contour plot of the ratio of Oort Cloud origin probability to scattered disk origin probability.  When generating this plot we only consider orbits with inclinations between 10$^{\circ}$ and 30$^{\circ}$.  The 'X' marks the orbit of 2006 SQ$_{372}$, and the triangle marks the orbit of 2000 OO$_{67}$  The dashed contour lines mark origin probability ratios of 10:1 and 1:1.}\label{fig:2}
\end{figure}

\begin{table*}[htbp]
\centering
\begin{tabular}{c c c c c c}
\hline
Active JFCs & Dormant: & New LPCs/yr & OC$_{{\rm in}}$:OC$_{{\rm out}}$ & 2000 OO$_{67}$ & 2006 SQ$_{372}$\\
 & active JFCs & & & ratio & ratio \\
(2 $< T_J <$ 3, & & ($q <$ 5 AU) & & (OC:SD) & (OC:SD) \\
$q <$ 2.5 AU) & & & & & \\
\hline
{\bf 250} & {\bf 2\footnotemark} & {\bf 1.5\footnotemark} & {\bf 3\footnotemark} & {\bf 14} & {\bf 16} \\
250 & 7\footnotemark & 1.5 & 1\footnotemark & 2.0 & 2.3 \\
250 & 7 & 1.5 & 3 & 5.8 & 6.9 \\
250 & 7 & 1.5 & 10\footnotemark & 19 & 23 \\
250 & 1\footnotemark & 10\footnotemark & 1 & 93 & 110 \\
250 & 1 & 10 & 3 & 270 & 320 \\
250 & 1 & 10 & 10 & 870 & 1100 \\
\hline
\end{tabular}
\caption{Parameters used in generating the Oort Cloud to scattered disk probability ratio and the resulting ratios.  Columns in order are: (1) the number of active JFCs in the current solar system, (2) the ratio of dormant to active JFCs in the current solar system, (3) the annual flux of dynamically new LPCs inside 5 AU, (4) the inner to outer Oort Cloud population ratio, (5) the derived origin probability ratio for 2000 OO$_{67}$, (6) the derived origin probability ratio for 2006 SQ$_{372}$.
}
\label{table:1}
\end{table*}

Our resulting probability ratio map is shown in Figure 2.  For this analysis, we use the normalization factors and inner Oort Cloud population that we have determined above.  In this figure, only orbital inclinations between 10$^{\circ}$ and 30$^{\circ}$ are included, as these most closely correspond to the true inclination of 2006 SQ$_{372}$.  It should be noted that the scattered disk contribution decreases with higher inclinations since it is concentrated toward the ecliptic much more than the Oort Cloud.  It can be seen in this figure that as semimajor axes approach 10$^3$ AU (the approximate value for SQ$_{372}$) the Oort Cloud origin probability increases by $\sim$2 orders of magnitude relative to the scattered disk.  This result supports the work of \citet{emel05} who conclude that the Oort Cloud should efficiently produce a population of planet-crossing orbits with $a \gtrsim 500$ AU beyond Saturn.

\footnotetext[2]{\citet{jew04}}
\footnotetext[3]{\citet{nes07}}
\footnotetext[4]{\citet{kaibquinn08}}
\footnotetext[5]{\citet{levdun97}}
\footnotetext[6]{\citet{dones04}}
\footnotetext[7]{\citet{bras06,bras08}}
\footnotetext[8]{\citet{fernmorb06}}
\footnotetext[9]{\citet{ever67a,ever67b}}

Thus, Figure 2 indicates that 2006 SQ$_{372}$ is 16 times more likely to originate from the Oort Cloud than from the scattered disk.  We must keep in mind, however, that the results shown in Figure 2 depend on a number of quite uncertain parameters.  Other combinations different from those used in Figure 2 are far from ruled out, and in Table 1 we list the resulting 2006 SQ$_{372}$ origin probability ratios for a range of other possible normalization factors and inner Oort Cloud populations.  It can be seen from this table that the probability of an Oort Cloud origin for 2006 SQ$_{372}$ remains higher than the scattered disk for any combination and may be over 1000 times the probability of a scattered disk origin.  From these results we believe that 2006 SQ$_{372}$ is the most distant LPC discovered to date.

\subsection{An Inner Oort Cloud Origin}

If 2006 SQ$_{372}$ is from the Oort Cloud, even more intriguing is that it almost certainly resided in the inner, unobserved regions of the Oort Cloud rather than the outer parts where all LPCs observed near Earth are thought to originate.  The dynamical reasoning is as follows.  Planetary perturbations must have been responsible for lowering the semimajor axis of this orbit to $\sim$10$^{3}$ AU, yet this random-walk process proceeds at a very slow pace.  From \citet{dun87}, it can be shown that 
\begin{equation}
\left|\frac{{\rm d}a}{{\rm d}t}\right|=\left({\rm 1 \times 10^6 yr}\right)^{-1}\times \left[\frac{D\left(q\right)}{{\rm 10^{-4} AU^{-1}}}\right]^2 \times\left(\frac{a^3}{{\rm 10^4 AU}}\right)^{1/2}
\end{equation}
and
\begin{eqnarray}
\nonumber \lefteqn{t_{D} = 2\left({\rm 1 \times 10^6 yr}\right) \times } \\
& & \left[\frac{{\rm 10^{-4} AU^{-1}}}{D\left(q\right)}\right]^2 \times \left[\left(\frac{{\rm 10^4 AU}}{a_{f}}\right)^{1/2} -\left(\frac{{\rm 10^4 AU}}{a_{0}}\right)^{1/2}\right]
\end{eqnarray}
where $D$ is the average energy kick per perihelion passage, $a_0$ is the original semimajor axis, $a_f$ is the final semimajor axis, and $t_D$ is the time required for the semimajor axis to diffuse down to $a_f$ from $a_0$ via planetary kicks.  Setting $a_f$ to 800 AU and $D$(25 AU) to 10$^{-4.7}$ AU$^{-1}$ \citep{dun87, fernbrun00} , we can then plot $t_D$ as a function of initial semimajor axis in the Oort Cloud in Figure 3.  As can be seen in this plot, $t_D$ is 130 Myrs for $a=10^4$ AU and increases for larger $a_0$.  

\begin{figure}[htbp]
\centering
\includegraphics[scale=.5]{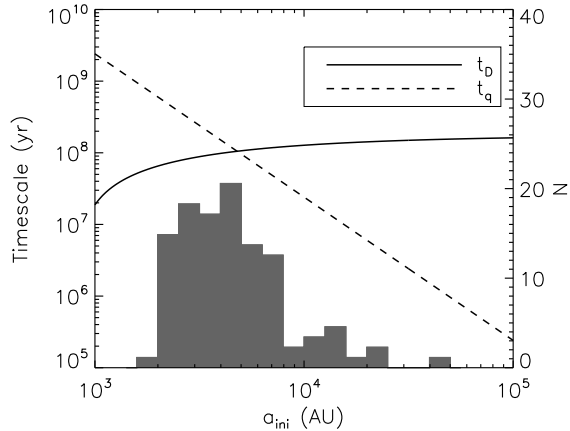}
\caption{Plot of the seimajor axis and perihelion diffusion timescales vs. initial semimajor axis.  The timescale for a perihelion shift of 10 AU (the shift necessary to detach SQ$_{372}$ from the planets) is shown by the dashed line, and the timescale for an orbital semimajor axis to random walk down to 800 AU is shown by the solid line.  In addition, we have overplotted a histogram of the initial Oort Cloud semimajor axes for all objects somewhat similar SQ$_{372}$ (20 AU $< q <$ 30 AU, 10$^\circ < i <$ 30$^\circ$, and $a <$ 800 AU) that we generated in our Oort Cloud simulations.}\label{fig:3}
\end{figure}

Left unchecked, the planets would be able to draw down the semimajor axes of any Oort Cloud comet within a couple hundred million years.  Before enough time can elapse for this to happen, however, most comets have their perihelia lifted back out of the planetary region again by the galactic tide.  If a comet's perihelion is moved outward by $\sim$10 AU from 25 AU to 35 AU, the planets cease to have a large effect on the dynamics, and the random-walk of semimajor axis stops.  The timescale for this to occur is given in \citet{dun87} as 

\begin{equation}
t_q = \left({\rm{1.3 \times 10^7}}{\rm{ yr}}\right)\left(\frac{\Delta q}{\rm{10 AU}}\right)\left(\frac{\rm{25 AU}}{q}\right)^{1/2}\left(\frac{\rm{10^4 AU}}{a}\right)^2
\end{equation}

Assuming a perihelion of 25 AU, $t_q$ is also plotted in Figure 3.  To produce SQ$_{372}$-like orbits from the Oort Cloud, $t_D$ must be shorter than $t_q$, otherwise $a=10^3$ AU will never be reached.  As can be seen in this plot, this is only the case for a subset of Oort Cloud bodies.  Only for initial semimajor axes below $\sim$5000 AU can the semimajor axis be drawn down to 10$^{3}$ AU before the perihelion is lifted out of the planetary region.  This process is verified when we examine the distribution of initial Oort Cloud semimajor axes for all of the SQ$_{372}$-like orbits we generate, which is also shown in Figure 3.  It can be seen in this figure that the distribution falls off quickly for  $a \gtrsim$ 10$^4$ AU.  Therefore, we can conclude from this argument that if 2006 SQ$_{372}$ is from the Oort Cloud, it is very likely from the inner 10$^4$ AU of the Oort Cloud.  

\subsection{Comparison with Other Objects}

\subsubsection{Comet C/2003 A2}
Comet C/2003 A2 \citep{gleas03,green03} is the only previously known LPC with perihelion beyond Saturn.  Because its perihelion is still within 2 AU of Saturn ($q = 11.4$ AU), its dynamics differ significantly from 2006 SQ$_{372}$.  This object's proximity to the strong energy kicks of Saturn make it just as likely to come from the outer Oort Cloud as the inner Oort Cloud.  This is because the timescale for Saturn to modify the semimajor axis of a comet is much shorter than Uranus or Neptune.  As a result, our argument for the inner Oort Cloud origin of 2006 SQ$_{372}$ cannot be applied to the orbit of Comet C/2003 A2, and we cannot constrain the region of the Oort Cloud where this comet previously resided.

\subsubsection{Sedna}
Given that we believe 2006 SQ$_{372}$ is from the inner Oort Cloud, a comparison with Sedna, another purported inner Oort Cloud body \citep{brown04}, is justified.  Sedna is thought to have been scattered on to its current orbit ($q = 76$ AU, $a = 487$ AU) by strong external perturbations of the Sun's birthplace environment early in the solar system's history \citep{morblev04,bras06,kaibquinn08}.  Because external forces of this magnitude no longer act in the Sun's vicinity, Sedna's orbit essentially ceased evolving Gyrs ago.  In contrast, because of a larger original semimajor axis, 2006 SQ$_{372}$ has recently had its perihelion pushed closer to Earth by the Galactic tide and passing stars.  Thus, 2006 SQ$_{372}$ represents a more proximate ($q \gtrsim$ 15 AU) population of inner Oort Cloud bodies whose orbits are still continually evolving.  

\subsubsection{2000 OO$_{67}$}
Out of all known TNOs, 2000 OO$_{67}$ (\cite{mil02,veil01}; $q = 20.7$ AU, $a = 552$ AU, $i = 20^\circ$) has an orbit most similar to 2006 SQ$_{372}$.  When we use Figure 2 to determine the probability that it is from the Oort Cloud, we find that the Oort Cloud should produce orbits similar to 2000 OO$_{67}$ at a rate that is at least 14 times that of the scattered disk, assuming the comet population numbers quoted in section 4.2.  Furthermore, because its perihelion is far beyond the orbit of Saturn, similar timescale arguments can be used to show that 2000 OO$_{67}$ must also come from the inner $\sim$10$^4$ AU of the Oort Cloud.  Thus, as suggested in \citet{emel05}, it appears that 2000 OO$_{67}$ is also a comet from the inner Oort Cloud.

\section{Summary}

We have discovered an object (2006 SQ$_{372}$) on an unusual orbit with a perihelion of 24.2 AU and a semimajor axis of 796 AU.  Because of the ability of the scattered disk and Oort Cloud to produce planet-crossing orbits in the outer Solar System we believe this object previously resided in one of these regions.  To determine which scenario is more probable we have simulated the production of similar orbits from both the scattered disk and the Oort Cloud.  The results of our simulations indicate that similar objects are produced from the Oort Cloud anywhere from 2.3 to 1100 times more often than the scattered disk, depending on the the populations of both regions.  We argue that a production rate ratio of 16 or higher is most likely and that an Oort Cloud origin for 2006 SQ$_{372}$ is therefore most probable.  Intriguingly, an Oort Cloud origin for 2006 SQ$_{372}$ implies that is very likely from the inner 10$^4$ AU of the Oort Cloud.  Otherwise, its perihelion would have been removed from the planetary region before its semimajor axis could be drawn down to $10^3$ AU.  This line of reasoning can also be applied to show that 2000 OO$_{67}$ must have had a similar history.  An inner Oort Cloud origin makes 2006 SQ$_{372}$ and 2000 OO$_{67}$ unique with respect to all known LPCs.  Furthermore, these bodies are members of an evolving, more proximate population of inner Oort Cloud bodies that are easier to detect because of their small perihelia compared to Sedna and other members of the extended scattered disk.  

Although we have shown 2006 SQ$_{372}$ is a probable LPC, this object cannot provide much new information about the structure of the Oort Cloud by itself.  The compilation of a large sample of similar Oort Cloud objects passing through the outer planetary region can, however, place better constraints on both the size and structure of the Oort Cloud.  The prospects for compiling such a sample in the near future are encouraging.  2006 SQ$_{372}$ was discovered by imaging the same fields of sky over multiple nights separated by periods of months during the SDSS-II SN survey.  In terms of both limiting magnitude and sky coverage, this survey is modest compared to the large synoptic surveys such as Pan-STARRS \citep{kais02,jew03} and LSST \citep{ivez08} being planned for the coming decade.  Thus, we can expect many objects similar to 2006 SQ$_{372}$ to be discovered in the near future.  Used in conjuction with dynamical modeling, these discoveries will provide many clues about the current size and structure of the Oort Cloud as well as the dynamical history of the solar system.

\section{Acknowledgements}

We would like to thank the reviewer, Alessandro Morbidelli, for insightful comments and suggestions that greatly improved the quality of this work.

This research was partially funded by a NASA Earth and Space Science Fellowship.  Most of our computing work was performed using the Purdue Teragrid computing facilities managed with Condor scheduling software (see http://www.cs.wisc.edu/condor).

Funding for the SDSS and SDSS-II has been provided by the Alfred P. Sloan Foundation, the Participating Institutions, the National Science Foundation, the U.S. Department of Energy, the National Aeronautics and Space Administration, the Japanese Monbukagakusho, the Max Planck Society, and the Higher Education Funding Council for England. The SDSS Web Site is http://www.sdss.org/.

The SDSS is managed by the Astrophysical Research Consortium for the Participating Institutions. The Participating Institutions are the American Museum of Natural History, Astrophysical Institute Potsdam, University of Basel, University of Cambridge, Case Western Reserve University, University of Chicago, Drexel University, Fermilab, the Institute for Advanced Study, the Japan Participation Group, Johns Hopkins University, the Joint Institute for Nuclear Astrophysics, the Kavli Institute for Particle Astrophysics and Cosmology, the Korean Scientist Group, the Chinese Academy of Sciences (LAMOST), Los Alamos National Laboratory, the Max-Planck-Institute for Astronomy (MPIA), the Max-Planck-Institute for Astrophysics (MPA), New Mexico State University, Ohio State University, University of Pittsburgh, University of Portsmouth, Princeton University, the United States Naval Observatory, and the University of Washington.

\nocite{york00,gun98,gun06,fuk96,frie08}

\bibliography{SQ372}

\end{document}